\newcounter{myctr}
\def\myitem{\refstepcounter{myctr}\bibfont\noindent\ifnum\themyctr>9\else\phantom{0}\fi\hangindent17pt\themyctr.\enskip}

\documentclass{ws-ijqi}
 
\begin{document}

\markboth{C.~Benedetti, F.~Buscemi, P.~Bordone and M.G.A.~Paris}
{Effects of classical environmental noise on entanglement dynamics}


\title{EFFECTS OF CLASSICAL ENVIRONMENTAL NOISE \\ 
ON ENTANGLEMENT AND QUANTUM DISCORD DYNAMICS}

\author{CLAUDIA BENEDETTI}
\address{Dipartimento di Fisica, Universit\`a degli Studi di Milano\\
Via Celoria 16, I-20133 Milano, Italy \\
claudia.benedetti@unimi.it}

\author{FABRIZIO BUSCEMI}
\address{ ARCES, Universit\`a di Bologna, Via Toffano 2/2, 40125 Bologna, Italy\\
fabrizio.buscemi@unimore.it}

\author{PAOLO BORDONE}
\address{Dipartimento di Scienze Fisiche, Informatiche e Matematiche, \\
Universit\`a di Modena e Reggio Emilia, \\ and Centro S3, CNR-Istituto di
Nanoscienze, Via Campi 213/A, 41125 Modena, Italy\\
paolo.bordone@unimore.it}
 
\author{MATTEO G.A. PARIS}
\address{Dipartimento di Fisica, Universit\`a degli Studi di Milano 
I-20133 Milano, Italy\\
CNISM, Unit\'a Milano Statale, I-20133 Milano, Italy\\
matteo.paris@fisica.unimi.it}

\maketitle

\begin{history}
\received{Day Month Year} 
\revised{Day Month Year}
\end{history}

\begin{abstract}
We address the effect of classical noise on the dynamics of quantum
correlations, entanglement and quantum discord, of two non-interacting
qubits initially prepared  in a Bell state.  The effect of noise is
modeled by randomizing the single-qubit transition amplitudes.  We
address both static and dynamic environmental noise corresponding to
interaction with separate common baths in either Markovian and
non-Markovian regimes. In the Markov regime, a monotone decay of the
quantum correlations is found, whereas for non-Markovian noise sudden
death and revival phenomena may occur, depending on the
characteristics of the noise. Entanglement and quantum discord show the
same qualitative behavior for all kind of noises considered. On the
other hand, we find that separate and common environments may play
opposite roles in preserving quantum correlations, depending on the
noise regime considered.
\end{abstract}

\keywords{Entanglement, quantum discord, classical noise}

\section{Introduction}
Quantum entanglement is a fundamental resource for quantum information processing, 
communication and exponential speed-up of some computational 
tasks~\cite{bennet,shor}. As a consequence, in the last years, there has been an
increasing interest in quantum correlations in various physical fields, ranging
from quantum optics \cite{cvqc1,cvqc2,vasile,cvqc3} 
to nanophysics~\cite{buscemi2010,buscemi2011}.
Furthemore, recently, it was pointed out that nonclassical correlations exist
which are more general, and possibly more fundamental then 
entanglement~\cite{ollivier,vedral,rev1}, 
the so-called quantum discord (QD), which is 
defined as the difference between total and classical correlations
in a system. Under suitable conditions QD has been proved to be more
robust than entanglement with respect to decoherence~\cite{maziero,werlag}.
Decoherence is indeed the main threat to the possibility to fully 
exploit the potentialities of quantum correlations for the above mentioned tasks.
In fact, the unavoidable interaction of systems with their environments induces 
loss of coherence, making the quantum parallelism, essential for quantum
computation, ineffective. 
On the other hand, environment can even resume quantum correlations or preserve them.
Thus, it is very important to analyze the effect of the various kinds of 
environmental noise on the entanglement and QD dynamics in realistic quantum systems, 
which can exhibit peculiar phenomena, such as entanglement~\cite{yu} and QD revival.
Markov noise is ascribed to environments with short, or rather instantaneous, 
self-correlations~\cite{yu2}. Non-Markovian noise~\cite{bellomo} is associated 
to environments with memory and may lead to the non-monotonic time dependence of  
entanglement and QD. 
Indeed, revival phenomena are found both for couples of qubits interacting directly 
or indirectly in a common quantum reservoir~\cite{ficek,mazzola} and for noninteracting 
qubits in independent non-Markovian quantum environments~\cite{bellomo}. Non-Markovianity
is also studied as a resource for quantum technology~\cite{vasile,chin}.

The effect of the environmental quantum noise on the entanglement dynamics between 
quantum systems has been interpreted in terms of the transfer of the correlations back 
and forth from the two-qubit system to the various parts of the global system. 
However, recent works showed that the non-monotonic time dependence
of the amount of quantum correlations may occur in two-qubit systems under the 
local action of a system-unaffected environment, such as classical random external 
potentials~\cite{zhou,lofranco}. Due to the classical nature of the
noise, in this case no back-action-induced correlations can be transferred 
from the system to the environment. Thus, the occurrence of entanglement revival in a 
non-Markovian classical environment is in contrast to the well-established interpretation
of revivals in terms of system-environment quantum back action, and rises the 
fundamental question of how one could explain the effect of classical noise on 
quantum correlations dynamics in bipartite systems.

In this work, we intend to analyze the role played by a classical noisy environment 
into the entanglement and QD dynamics between two quantum particles in a simple system, 
which not only allows us to relate the intrinsic features of the two-particle dynamics 
to the revival of the correlations, but can also be of great relevance in
various physical phenomena. Specifically, we consider two qubits non interacting between
each other and coupled to noise in different and common environments. Two different
kind of noise are considered: static and random telegraph noise (RTN).
The former has also been used to describe electron transport~\cite{lahini} and 
photon propagation\cite{Thom} in disordered structures.
The latter represents one of the common environmental noises affecting charge
carriers in nanodevices and also plays a key role into the building up of
$1/f^\alpha$ noises appearing in a large number of solid-state systems\cite{Fuj,Kur,paladino2,paladino3}.
In this paper, we used an analytical approach to solve the two-qubit model,
and therefore to estimate its time evolution. Numerical techniques are also
adopted to evaluate the dynamics of quantum correlations. Static noise is here
used to simulate a non-Markovian environment. To this aim random time-independent terms
are inserted in the off-diagonal coefficients of the single-qubit Hamiltonians. 
On the other hand, a dynamic disorder can model both a non-Markov environment, 
expressed by a slow RTN, and a Markovian noise, in the limit of fast RTN.
In both cases single-qubits transition amplitudes are time-dependent and are assumed
to stochastically switch between two values.

The paper is organized as follows. The physical model adopted is described
in Sec. 2. In Sec. 3 are presented the results for the various cases considered:
static and dynamic noise, different and common environments, Markov and non-Markov
regime. Conclusions and outlooks are given in Sec. 4.

\section{The Physical Model}
In this section we describe a model of two qubits, initially entangled, subject
to a noisy classical environment. A static and a random telegraph noise 
are accounted for in different conditions, specifically local and non-local 
interactions between qubits and environments are considered.
Furthermore the estimators of quantum correlations are described.

The dynamics of the system is ruled by the Hamiltonian:
\begin{equation}  \label{Ham1}
H(t)=H_A(t)\otimes\mathbb{I}_B+\mathbb {I}_A\otimes H_B(t)\ ,
\end{equation}
where $H_{A(B)}(t)$ is the single-qubit Hamiltonian defined as:
\begin{equation}  \label{Ham2}
H_{A(B)}(t)=\varepsilon\ \mathbb{I}_{A(B)}+\nu c_{A(B)}(t){\sigma_x}_{A(B)}.
\end{equation}
$\mathbb{I}_{A(B)}$ and  ${\sigma_x}_{A(B)}$ are the identity operator 
and the Pauli matrix of the subspace of the qubit $A(B)$, respectively. 
$\varepsilon$ is the  qubit  energy in the absence of noise (here we 
assume  energy degeneracy), $\nu$ is the system-environment  
coupling constant and $c_{A(B)}(t)$ is a random parameter related to the specific
characteristics of the noise. This  model has already been  used to describe
a quantum walk of two not-interacting particles in a  noisy lattice~\cite{Bordo}.

The Hamiltonian given in Eq.~(\ref{Ham1}) is stochastic due to 
the randomness of the parameter $c(t)$, thus leading to a stochastic
 time evolution of the quantum states.  Once a choice of the 
noise parameter is performed, the corresponding  evolution operator
is given by $U(t)=e^{-i \int H(t) dt}$ (hereafter $\hbar$=1).  When the latter is applied
to the initial  state the specific system dynamics
is obtained. Finally, the density matrix describing the two qubits
 is evaluated by performing an average over the different noise configurations. 
The two qubits are initially  prepared in  the Bell state $|\Phi^+\rangle$=$1/\sqrt{2} \left( |0_A0_B \rangle+ |1_A1_B \rangle\right)$.

 \subsection{Static noise}\label{statnoisemodel}
 The static noise has already been used to study the propagation
 of quantum particles in disordered systems, specifically in 
 optical coupled waveguides and quantum walks~\cite{Thom,Bordo}.
 In agreement with these works,  to model the static noise  the adimensional parameters $c(t)$ are assumed to be time-independent
 random variables following the flat probability distribution given by  $P(c)$=$1/\Delta_c$ for $|c-c_0|\leqslant \Delta_c/2$ 
 and 0 otherwise. $c_0$ denotes the mean value of the distribution and $\Delta_c$ quantifies
 the disorder of the environment.  The autocorrelation function of $c$   reads $\langle  \delta c(t) \delta c(0)\rangle $=$\Delta_c^2/12$,
 and therefore its power spectrum is given by a $\delta$-function centered on zero frequency. This means that the memory
 effects of the static noise do not  vanish at any time. As a consequence, we can classify such a noise as non-Markovian.
 
The local coupling between qubits and environment  has  been mimicked 
by assuming that the noise parameters of  the two qubits $c_A$ and $c_B$
are uncorrelated and described by two flat probability distributions 
each characterized by the same  $c_0$  and $\Delta_c$. On the other hand,
for the case of two qubits  interacting with  a common environment
we set  $c_A$=$c_B$, which means that  the same  random noise parameter
is used in  both  single-qubit Hamiltonians of  Eq.~(\ref{Ham2}) to model the noise effects.
 
As stated above, to describe the  full dynamics of the two-qubit
system subject to the disordered environment  the average over
all the possible noise configurations is required.  For the static noise,
such an average is given by  the integral  of the time-evolved states 
each corresponding to a specific choice of  the noise parameters.
In other words, for the case of local coupling to different  environments 
the two-qubit density matrix  $\rho_{de}(t)$, at time $t$, reads
 \begin{equation}  \label{rhode}
 \rho_{de}(t)=\int_{c_0-\frac{\Delta_c}{2}}^{c_0+\frac{\Delta_c}{2}} \int_{c_0-\frac{\Delta_c}{2}}^{c_0+\frac{\Delta_c}{2}} \!\!\!dc_A dc_B \:P(c_A) P(c_B) \rho(c_A,c_B,t)
\end{equation}
 where $\rho(c_A,c_B,t)$=$U(c_A,c_B,t)|\Phi^+\rangle \langle \Phi^+ | U^{\dag}(c_A,c_B,t)$. When the two  qubits are coupled to a common environment, 
the mixed state   of the system $\rho_{ce}(t)$ is given by 
  \begin{equation}  \label{rhoce}
 \rho_{ce}(t)=\int_{c_0-\frac{\Delta_c}{2}}^{c_0+\frac{\Delta_c}{2}} \!\!\!\! dc\: P(c) \rho(c,t)
\end{equation} 
 where $\rho(c,t)$=$U(c,t)|\Phi^+\rangle \langle \Phi^+ | U^{\dag}(c,t)$. 
The explicit form of $ \rho_{de}(t)$ and $ \rho_{ce}(t)$ is presented in
Sec.~\ref{resu}.

 \subsection{Random telegraph noise}\label{rtnsec}
 The other kind of noise we examine is the RTN,
 which is able to describe a number of   typical phenomena affecting
 solid-state devices on the nanoscale~\cite{Fuj,Kur}. Here $c(t)$ is assumed
 to flip randomly between the values -1 and 1 at rate $\gamma$. The autocorrelation
 function falls off exponentially as $\langle  \delta c(t) \delta c(0)\rangle $=$e^{-2\gamma t}$
 and its power spectrum exhibits the well-known Lorentzian shape $4\gamma/(\omega^2+4\gamma^2)$.
 Two different regimes of the decay of quantum correlations are identified,  according to 
 the ratio between the system-environment coupling $\nu$ and the switching rate $\gamma$ of RTN~\cite{zhou}.
 For $ \nu/\gamma \ll 1$  the  Markovian  regime is found, while for $ \nu/\gamma \gg 1$
 the  non-Markovian behavior is obtained.

As discussed by Abel~\emph{et al.}~\cite{Abel}, the dichotomic stochastic behavior  of $c(t)$ induces a random phase factor 
$\varphi_{A(B)}(t)=-\nu \int_0^t dt^{\prime} c_{A(B)}(t^{\prime} )$
in the evolution of the single-qubit states. The two-qubit density matrix
is given by the average over the random phase factor, namely 
 \begin{equation}  
\rho_{de}(t)=\left \langle  \left \langle \rho(\varphi_A(t),\varphi_B(t)) \right \rangle_{\varphi_A} \right \rangle_{\varphi_B}\label{rtnde}
\end{equation}
for local coupling qubit-environment, and 
 \begin{equation}  
\rho_{ce}(t)  =\left \langle \rho(\varphi(t)) \right \rangle_{\varphi}\label{rtnce}
\end{equation}
for  coupling of the two qubits with a common environment.  Also in this case,
the explicit forms  of $ \rho_{de}(t)$ and $ \rho_{ce}(t)$ are given in
Sec.~\ref{resu}.

 \subsection{Estimators of quantum correlations}
Here, we briefly review the estimators used
to quantify entanglement and  quantum discord in our system.

Entanglement  is evaluated in terms of negativity~\cite{negat}
defined as:
\begin{equation} \label{Negati}
 \mathcal{N}=2\left|\sum_i\lambda_i^-\right|,
\end{equation}
where $\lambda_i^-$ are the negative eigenvalues of the partial transpose
of the density matrix of the bipartite system. Negativity ranges from
zero, for separable states, to one, for maximally entangled states.

QD is the difference between the total and the classical correlations in a system
so representing its  degree of quantumness~\cite{ollivier}. It is defined as
\begin{equation}
\mathcal{Q}=\mathcal{I}-\mathcal{C},
\end{equation}
where $\mathcal{I}$ is  the quantum mutual information 
which gives a measure of the total amount of correlations of the system.
It is defined as: 
\begin{equation}
\mathcal{I}=S(\rho^A) +S(\rho^B)-S(\rho),
\end{equation}
where  $S$ is the von Neumann entropy  and $\rho^{A(B)}$ indicates
the reduced density matrix of the subsystem $A(B)$.
$\mathcal{C}$ denotes  the measurement-induced quantum mutual information,
namely the classical correlations. The latter reads
\begin{equation}
\mathcal{C}=\max_{\{B_k\}}\{ S(\rho_A)-S(\rho|\{B_k\})\},
\end{equation}
with $S(\rho|\{B_k\})$ indicating  the quantum conditional entropy with respect to the set of measurements $\{B_k\}$ performed locally  on subsystem $B$.
Usually to compute QD is not an easy task, since it involves a maximization procedure.  But in the case of two-qubit X states an analytical expression for $\mathcal{Q}$ was derived by Luo~\cite{luo}.

\section{Results}\label{resu}
In this Section, we present the explicit forms, as a function of time,
of the various mixed states of  the two qubits, and 
the time behavior of entanglement and QD
for  all the considered physical configurations.
Explicit derivation for the average density matrices is shown in  \ref{appendice}.
\subsection{Static noise}
When the two qubits are affected by static noise,
their  density matrices, obtained from Eqs.~(\ref{rhode}) and (\ref{rhoce}),  take the form:
\begin{equation} \label{rhostat}
 \rho_{de(ce)}(t)=\left(\begin{array}{cccc}
         \frac{1}{4}+\alpha_{de(ce)}&-\beta_{de(ce)}&-\beta_{de(ce)}&\frac{1}{4}+\alpha_{de(ce)}\\
\beta_{de(ce)}&\frac{1}{4}-\alpha_{de(ce)}&\frac{1}{4}-\alpha_{de(ce)}&\beta_{de(ce)}\\
\beta_{de(ce)}&\frac{1}{4}-\alpha_{de(ce)}&\frac{1}{4}-\alpha_{de(ce)}&\beta_{de(ce)}\\
\frac{1}{4}+\alpha_{de(ce)}&-\beta_{de(ce)}&-\beta_{de(ce)}&\frac{1}{4}+\alpha_{de(ce)}
         \end{array}\right),
\end{equation}
where 
\begin{eqnarray} 
\alpha_{de}=\frac{1}{( 2 \Delta_c \nu t)^2} \cos{(4c_0 \nu t)} \sin^2{(  \Delta_c \nu t)}\quad\textrm{and}\quad \beta_{de}=\frac{i}{( 2 \Delta_c \nu t)^2} \sin{(4c_0 \nu t)} \sin^2{(  \Delta_c\nu t)} \nonumber \\ 
\alpha_{ce}=\frac{1}{8 \Delta_c \nu t} \cos{(4c_0 \nu t)} \sin{(  2\Delta_c \nu t)}\quad\textrm{and}\quad \beta_{ce}=-\frac{i}{8 \Delta_c \nu t}\sin{(4c_0 \nu t)} \sin{(  2\Delta_c \nu t)} .\nonumber  
\end{eqnarray}
It should be noticed that, for  both local and non-local qubit-environment coupling,
at sufficiently long times,  the matrices take an X form where all non-vanishing
coefficients are equal. This means that the asymptotic two-qubit state is given
by a statistical mixture of two Bell states, specifically 
$1/2\left(|\Phi^+ \rangle  \langle\Phi^+ |+|\Psi^+ \rangle  \langle\Psi^+|\right)$.

\begin{figure}[htpb]
   \begin{center}
   \includegraphics*[width=0.475\textwidth]{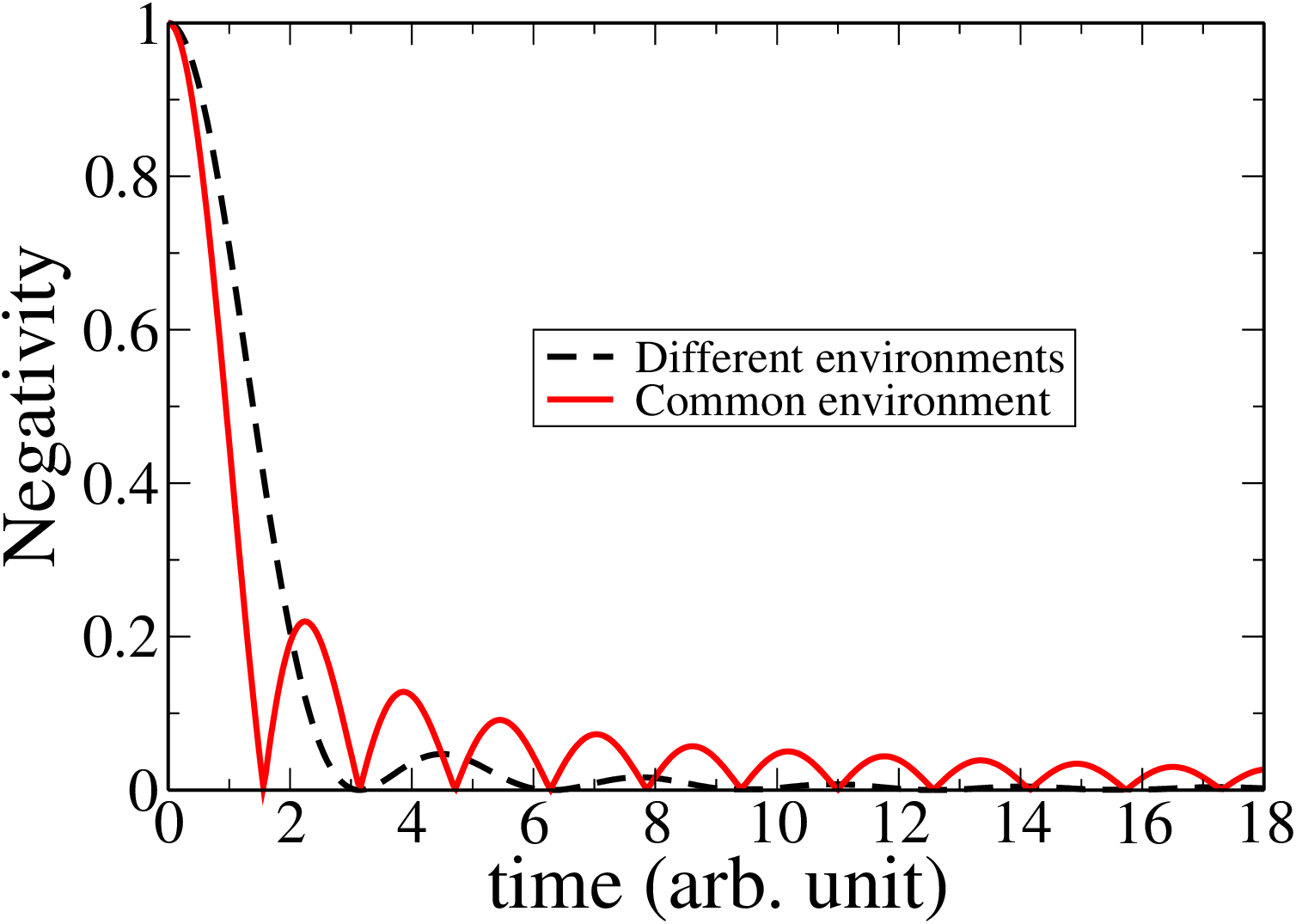}
\includegraphics*[width=0.475\textwidth]{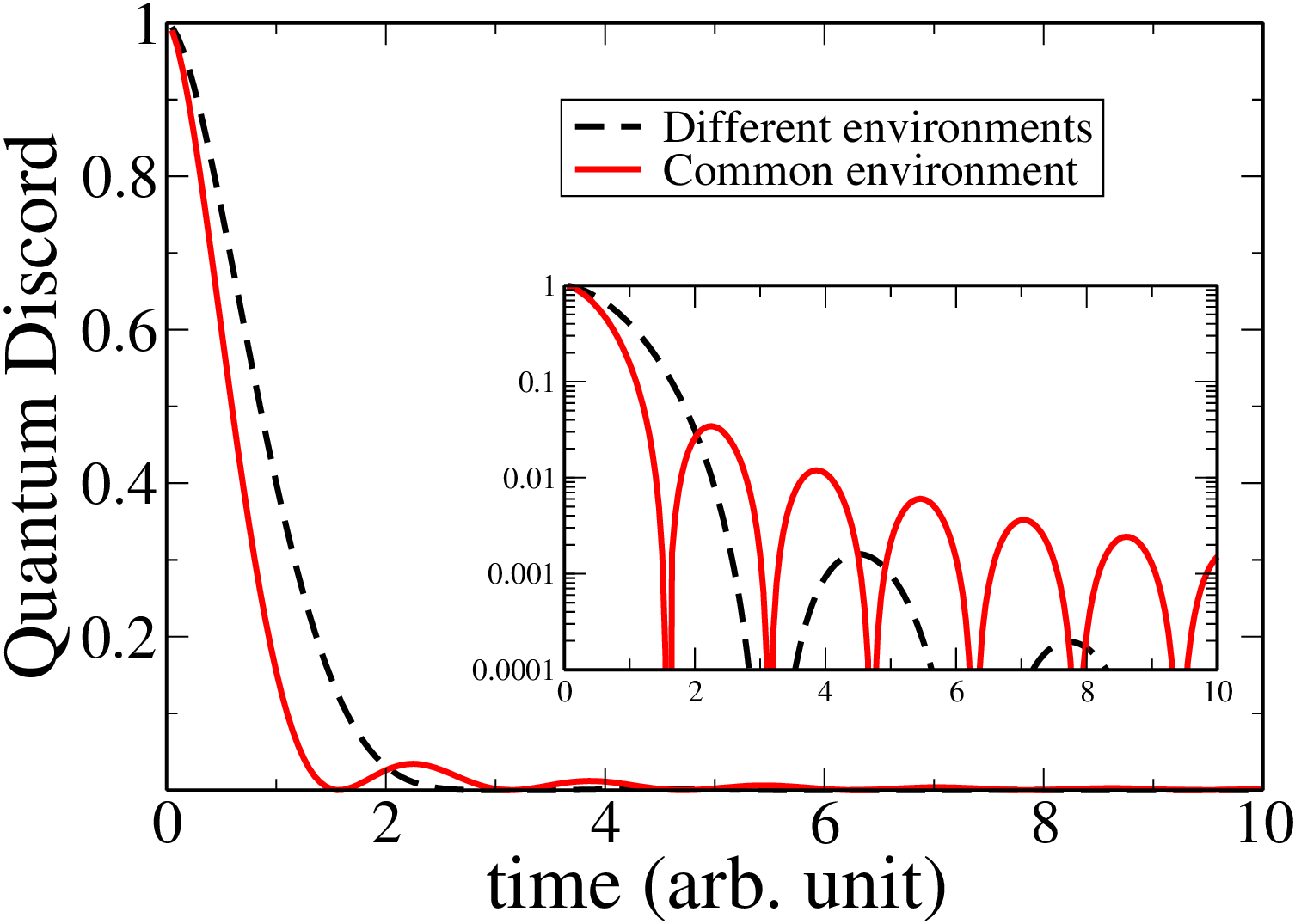}
     \caption{ \label{fig1} Time evolution  of  negativity (left panel) and QD (right panel) for two qubits subject to a static noise 
     for different  and common environments. To highlight  the revivals of QD, its behavior 
     is reported also in logarithmic scale in the inset of the right panel. }
   \end{center}
 \end{figure}

Performing the calculation starting from the definition given  in Eq.~(\ref{Negati}),
the negativity can be be expressed as
\begin{equation} 
 \mathcal{N}_{de}(t)=\left(\frac{\sin{(\Delta_c\nu t)}}{\Delta_c\nu t}\right)^2\quad\textrm{and}\quad  \mathcal{N}_{ce}(t)=\frac{\sin{(2\Delta_c\nu t)}}{2\Delta_c\nu t}.
\end{equation}
Such a result clearly shows  a non-monotonic time decay of entanglement
as expected from the non-Markovian nature of the noise. Peculiar phenomena as
sudden death and revival are observed
(see  left panel of Fig.~\ref{fig1}). Furthermore, we find that in the case of common environment 
entanglement is more robust with respect to  the case of  different environments. Indeed, when the two
qubits are coupled with a common environment, the latter can be interpreted as a sort of 
interaction mediator between the qubits themselves. Such an interaction somehow contributes 
to build up quantum correlations even if decohering effects of the environment are still 
dominant and lead to a power-like decaying profile of entanglement.  
An analogous behavior is shown by QD, as displayed in 
 the right panel of Fig.~\ref{fig1}. Note that, unlike negativity, QD has been evaluated numerically
 by means of a maximization algorithm.

\subsection{Random telegraph noise}
For the case of a RTN affecting the dynamics of two qubits,
the time evolution of the mixed state is evaluated
from the average over the random phase factor $\varphi(t)$. 
After a straightforward calculation,
we obtain
\begin{equation} \label{rhortn}
 \rho_{de(ce)}(t)=   \frac{1}{4}\left(\begin{array}{cccc}
      1+\Lambda_{de(ce)}& 0& 0&1+\Lambda_{de(ce)}\\
0&1-\Lambda_{de(ce)}&1-\Lambda_{de(ce)}&0\\
0&1-\Lambda_{de(ce)}&1-\Lambda_{de(ce)}&0\\
 1+\Lambda_{de(ce)}& 0& 0&1+\Lambda_{de(ce)}
         \end{array}\right),
\end{equation}
with $\Lambda_{de}$=$D_{2\nu}^2(t)$ and  $\Lambda_{ce}$=$D_{4\nu}(t)$.
As shown elsewhere~\cite{Abel}, the function $D(t)$  corresponds to the average phase factor
$\langle e^{i\varphi(t)}\rangle$ and can be expressed as
\begin{equation} \label{ddtdif}
D_{m\nu}(t)=\left\{ \begin{array}{ccc}e^{-\gamma t}\left[\cosh {\left(\delta_{m\nu}t\right)} +\frac{\gamma}{\delta_{m\nu}}\sinh {\left(\delta_{m\nu}t\right)}\right] &\quad \textrm{for} \quad& \gamma > m\nu  \\  e^{-\gamma t}\left[\cos {\left(\delta_{m\nu}t\right)} +\frac{\gamma}{\delta_{m\nu}}\sin {\left(\delta_{m\nu}t\right)}\right] &\quad \textrm{for} \quad &\gamma < m\nu
\end{array}\right . ,
  \end{equation}
where $\delta_{m\nu}$=$\sqrt{|\gamma^2 -(m\nu)^2|}$ with $m\in\{2,4\}$.
Unlike the static noise, the two-qubit density matrix assumes
an X form both for different and common environment at any time $t$.  It is worth noting
that, for both RTN configurations, the long time asymptotic form of the states obtained
is the same found for the static noise. Using again the definition of 
Eq.~(\ref{Negati}), negativity reads:
\begin{equation} 
\mathcal{N}_{de}(t)=D_{2\nu}^2(t)\quad\textrm{and}\quad 
\mathcal{N}_{ce}(t)=|D_{4\nu}(t)|.
\end{equation}
In this case, due to the X form of the states, it is possible to give
the analytical expressions for QD by following the  Luo approach~\cite{luo}.
They read:
\begin{eqnarray} 
\mathcal{Q}_{de}(t)&=&\frac{1}{2}\left[(1+D_{2\nu}^2)\log_2{\left(1+D_{2\nu}^2\right)}+(1-D_{2\nu}^2)\log_2{\left(1-D_{2\nu}^2\right)}\right] \nonumber \\
\mathcal{Q}_{ce}(t)&=&\frac{1}{2}\left[(1+D_{4\nu})\log_2{\left(1+D_{4\nu}\right)}+(1-D_{4\nu})\log_2{\left(1-D_{4\nu}\right)}\right].
\end{eqnarray}

\begin{figure}[htpb]
   \begin{center}
   \includegraphics*[width=0.475\textwidth]{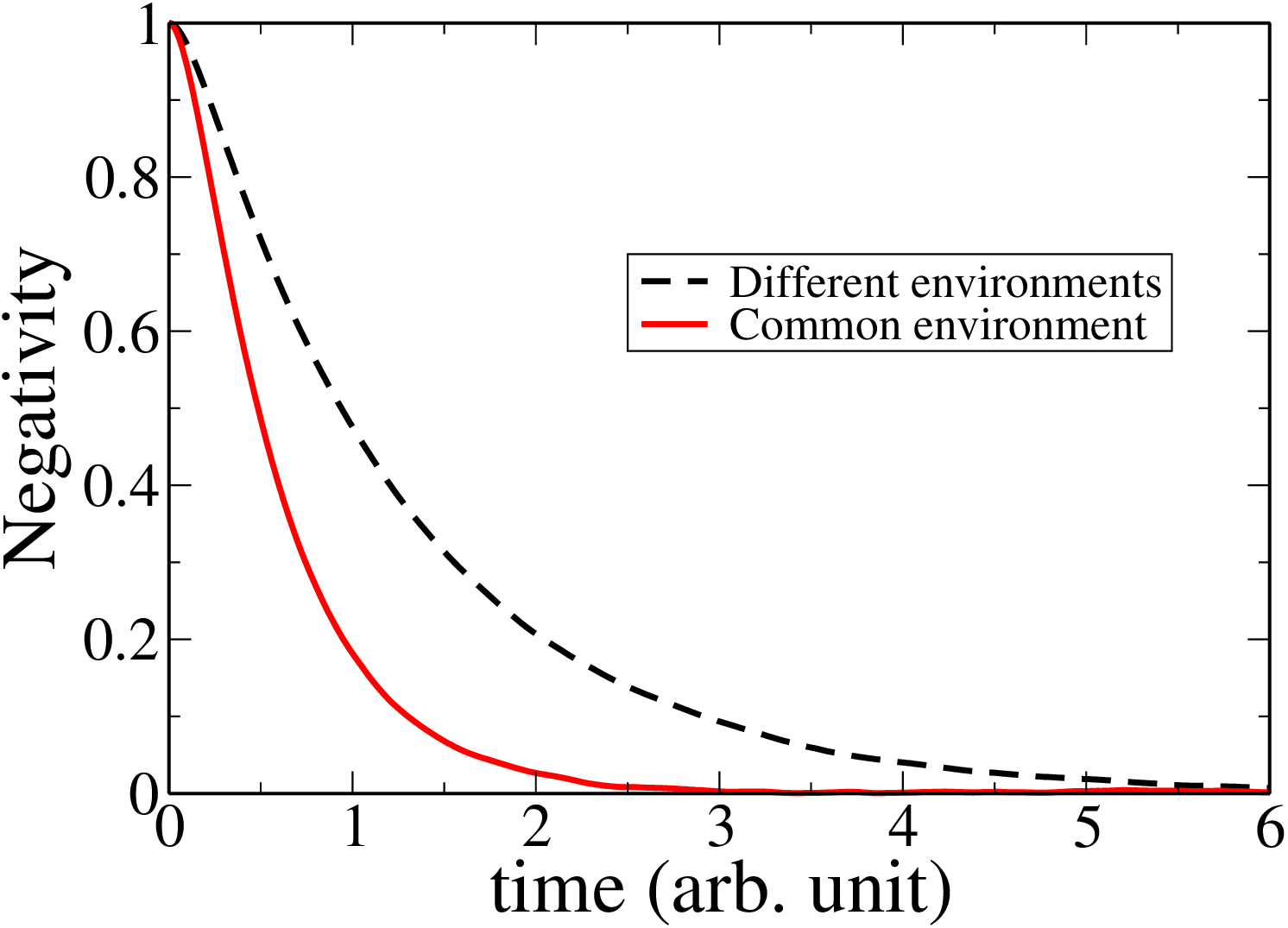}
\includegraphics*[width=0.475\textwidth]{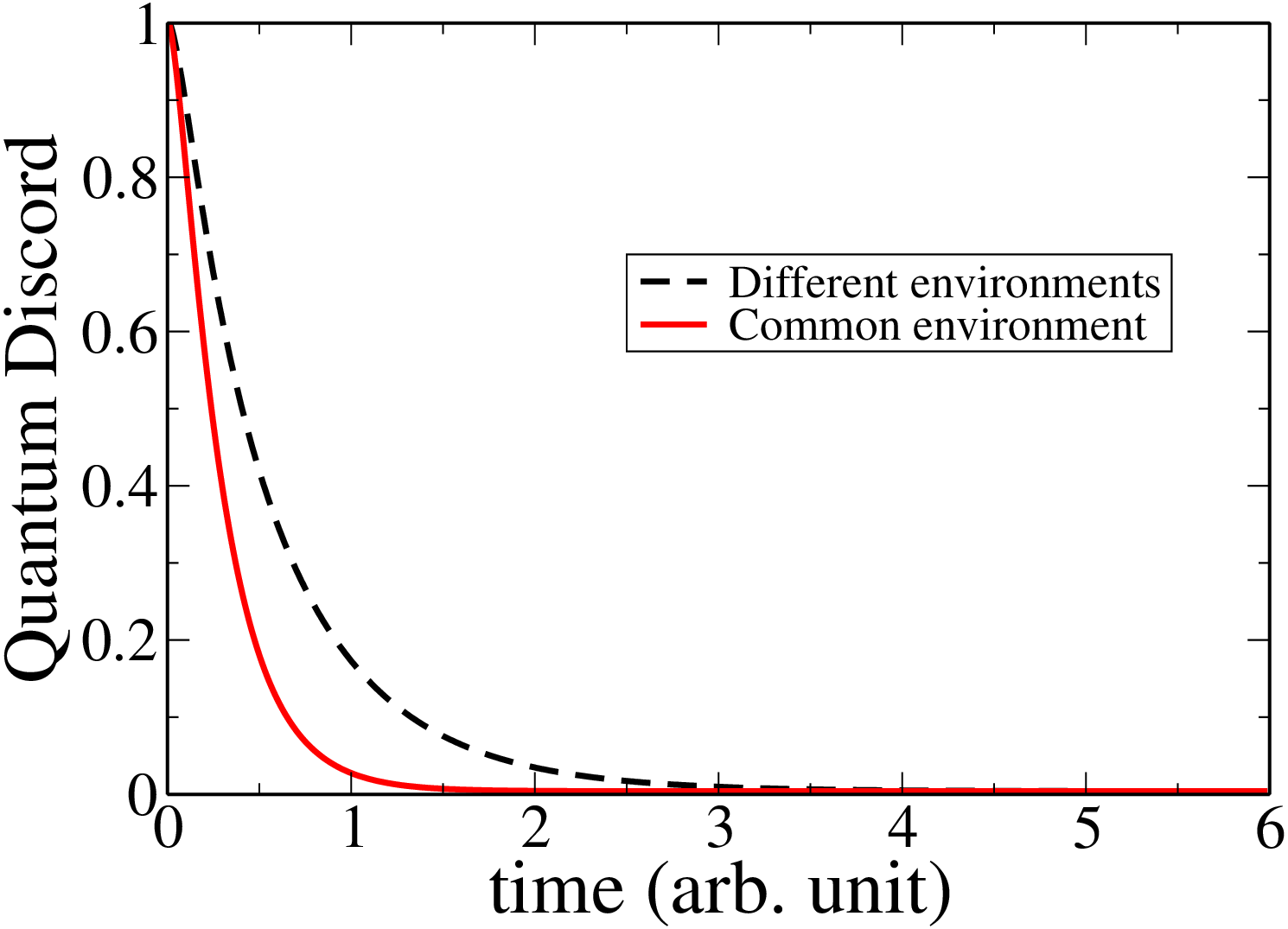}
 \includegraphics*[width=0.475\textwidth]{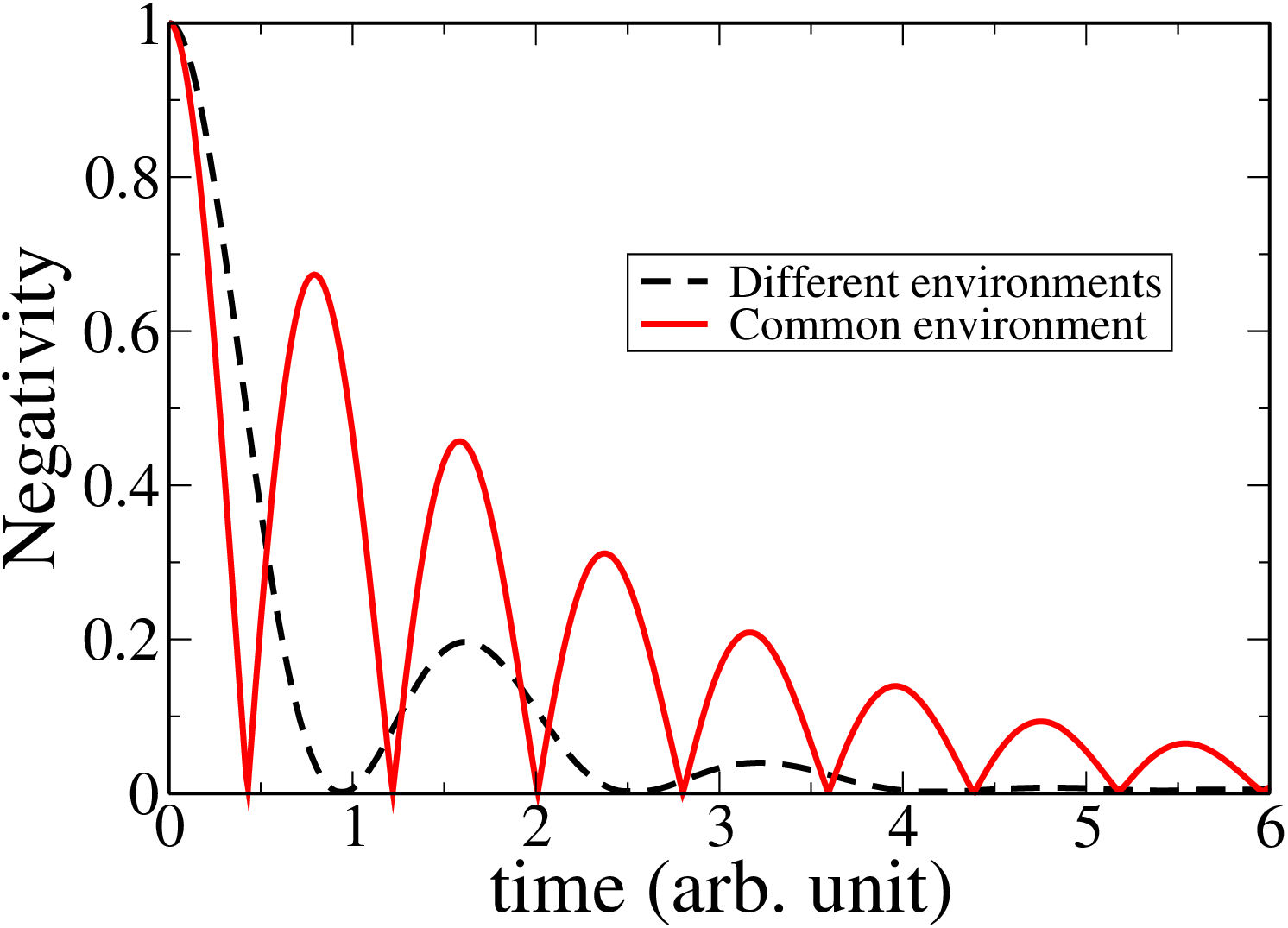}
\includegraphics*[width=0.475\textwidth]{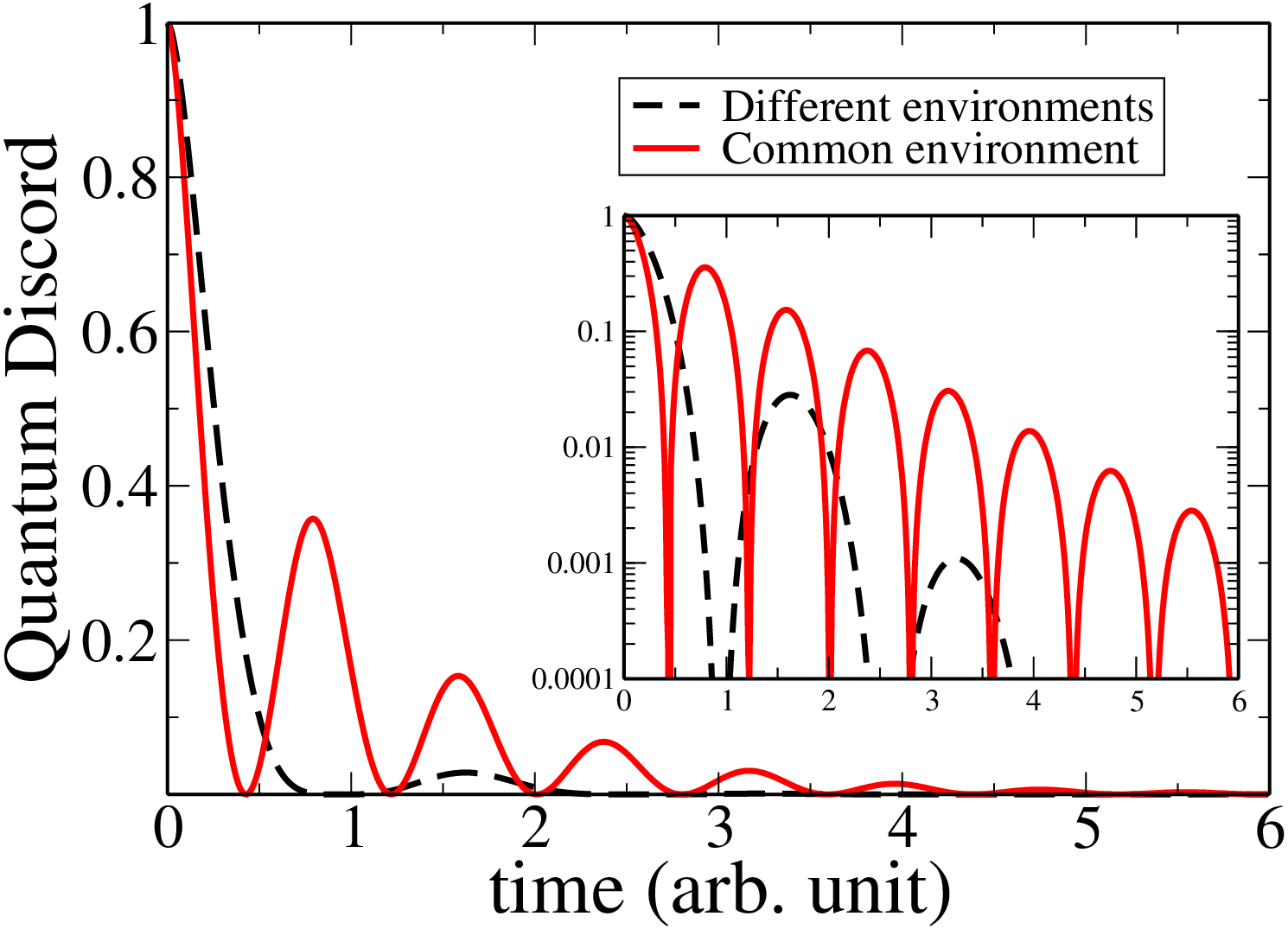}
     \caption{ \label{fig2} Top panels: Time evolution  of  negativity (left) and QD (right) for two qubits subject to a RTN
     for different  and common environments when $\nu/\gamma=0.2$ (Markovian regime). Bottom panels: 
     Time evolution  of  negativity (left) and QD (right) for two qubits subject to a RTN
     for different  and common environments when $\nu/\gamma=5$ (non-Markovian regime). 
To highlight  the revivals of QD, its behavior is reported also in logarithmic scale 
in the inset of the right panel.}
   \end{center}
 \end{figure}

The dynamics of $\mathcal{N}$ and $\mathcal{Q}$ is shown in Fig.~\ref{fig2}. 
In agreement with the outcomes of  previous works~\cite{Bordo},
in the Markovian regime both entanglement and QD exhibits an exponential decay with time. On the other hand,
in the non-Markovian regime the amount of quantum correlation are a damped oscillating function of time,
thus displaying sudden death and revival phenomena. Finally, it should be noticed that 
our calculations show that, in the Markovian regime, the quantum correlations for  the  case of coupling  
of two qubits with the common environment are weaker with respect to the case of local qubit-environment
interaction. Such  a behavior  is in contrast with the one found for non-Markovian noises
 and can be ascribed  to the fact that indirect qubit-qubit interaction is here more 
effective in destroying both entanglement and QD. 
 
\section{Conclusions}\label{conc}
In this paper we have analyzed the effects of both  independent and common
noisy environments on the dynamics of quantum correlations of two
non-interacting qubits.  Negativity and QD have been computed, using
both analytical and numerical techniques. Static and slow RTN are
classified as non-Markovian noises, while the fast RTN mimics a
Markovian environment.  We showed that, starting from a maximally
entangled state, the quantum correlations display different decaying
behaviors, depending on the nature of the considered noise.  In
particular, Markovian environments lead to a monotonic decay, while in
non-Markov regimes sudden death and revival phenomena are present, in 
agreement with previous results in the literature\cite{zhou,paladino}.
Specifically, here we ascribe the decay of revivals obtained in the static noise 
 to the continuous nature of the random noise parameter. In fact our analysis
clearly shows the suppression of quantum correlations at long times unlike previous investigations
where an environmental noise mimicked by means of  dichotomic time-independent noise parameter leads
to non decaying entanglement revivals\cite{lofranco}.\\
For
both static noise and slow RTN, our results highlight that a common
environment preserves better quantum correlations. It is worth noting
that the opposite is found for qubits subject to a fast RTN, where the
effect of a common noise results in a faster decay of correlations with
respect to the case of independent environments.
\par
An interesting generalization of our model is the possibility to sum up
many RTNs, each with a specific switching rate, in order to obtain
$1/f^{\alpha}$ noise.  This noise is ubiquitous in solid state
devices\cite{weissman} and can be simulated considering a linear
superposition of different RTNs, with an appropriate distribution of
their switching rates.  Furthermore, our study of the effects of static
and dynamic noise on the evolution of quantum correlations in two-qubit
models, has a natural and quite promising generalization in the case of
systems with a larger number of degrees of freedom.  For instance the
model here studied can be used to extend, to the presence of noise,
previous works on quantum walks on a one-dimensional
lattice\cite{benedettiq}.  In this case the environmental noise would be
a consequence of the lattice disorder.  
Work along these lines is in progress and results will be reported
elsewhere. 

\section*{Acknowledgements}
This work has been supported by MIUR project FIRB LiCHIS-RBFR10YQ3H
and by the Finnish Cultural Foundation.

\appendix
\section{Explicit derivation of the two-qubit density matrix}\label{appendice}
Given the Hamiltonian in Eqs. \eqref{Ham1} and \eqref{Ham2}, the associated evolution operator reads:
\begin{align}
 U_{de(ce)}=&e^{-2i\beta t}
\left(\begin{array}{cccc}
         A_{de(ce)} & -iB_{de(ce)}& -iC_{de(ce)} & -D_{de(ce)} \\
	  -iB_{de(ce)}&  A_{de(ce)}& -D_{de(ce)}& -iC_{de(ce)}\\
	   -iC_{de(ce)} & -D_{de(ce)}& A_{de(ce)} & -iB_{de(ce)}\\
	   -D_{de(ce)}&  -iC_{de(ce)}& -iB_{de(ce)}&  A_{de(ce)}
         \end{array}\right),
\end{align}
where
\begin{eqnarray}
&A_{de}=\cos(\varphi_A(t))\cos(\varphi_B(t))\quad\quad &A_{ce}= \cos^2(\varphi(t))\nonumber \\ 
&B_{de}=\cos(\varphi_A(t))\sin(\varphi_B(t))\quad\quad &B_{ce}=\cos(\varphi(t))\sin(\varphi(t))\nonumber  \\
&C_{de}=\sin(\varphi_A(t))\cos(\varphi_B(t))\quad\quad &C_{ce}=B_{ce}=\cos(\varphi(t))\sin(\varphi(t)) \nonumber \\ 
&D_{de}=\sin(\varphi_A(t))\sin(\varphi_B(t))\quad\quad &D_{ce}=\sin^2(\varphi(t))\nonumber ,
\end{eqnarray}
with the phase $\varphi(t)=ct$ in the case of static noise and 
$\varphi(t)=-\nu\int_0^t c(t')dt'$ for the RTN.
Applying the evolution  operator to the initial state $\rho_0=|\Phi^+\rangle\langle\Phi^+|$, it 
is possible to obtain the time-evolved density matrix of the system corresponding to a specific choice of the 
random parameter:  
\begin{align}
&\rho(\varphi,t)=U_{de(ce)}^+\rho_0U_{de(ce)}^{\dagger}=\nonumber\\
&=\frac{1}{2}
\left(\begin{array}{cccc}
           \tilde{A}_{de(ce)} &\tilde{B}_{de(ce)} &\tilde{B}_{de(ce)} & \tilde{A}_{de(ce)} \\
	    -\tilde{B}_{de(ce)}& \tilde{C}_{de(ce)}&\tilde{C}_{de(ce)} & -\tilde{B}_{de(ce)}\\
	   -\tilde{B}_{de(ce)} & \tilde{C}_{de(ce)}&\tilde{C}_{de(ce)} & -\tilde{B}_{de(ce)}\\
	    \tilde{A}_{de(ce)}&\tilde{B}_{de(ce)} &\tilde{B}_{de(ce)} & \tilde{A}_{de(ce)}
         \end{array}\right)\label{rhostat1}
\end{align}
where:
\begin{align}
&\tilde{A}_{de(ce)}=(A_{de(ce)} - D_{de(ce)})^2\nonumber\\
&\tilde{B}_{de(ce)}= i (B_{de(ce)} + C_{de(ce)} ) (A_{de(ce)}- D_{de(ce)})\label{coeffstat} \\ 
&\tilde{C}_{de(ce)}= (B_{de(ce)} + C_{de(ce)})^2\nonumber  
\end{align}
The average of the expression \ref{rhostat1} over the possible values of the noise parameters
gives the state of the system at time $t$ and corresponds to the integral  
\begin{equation}
 \langle\dots\rangle_{\varphi}=\int d\varphi(\dots)p(\varphi,t)
\end{equation}
where for the static noise the phase distribution $p(\varphi,t) =P(c)/t$ where $P(c)$ is the
flat distribution given in Sec. \ref{statnoisemodel}, while for the RTN,  it takes the form\cite{bergli}:
\begin{align}
 p(\varphi,t)&=\frac{1}{2}e^{-\gamma t}\left\{[\delta(\varphi+\nu t)+\delta(\varphi-\nu t)]+\frac{\gamma}{\nu}[\Theta(\varphi+\nu t)+\Theta(\varphi-\nu t)]\right\}\nonumber\\
&\times\left[\frac{I_1\left(\gamma t\sqrt{1-(\varphi/\nu t)^2}\right)}{\sqrt{1-(\varphi/\nu t)^2}}+I_0\left(\gamma t\sqrt{1-(\varphi/\nu t)^2}\right)\right].
\end{align}
$I_k(x)$ is the modified Bessel function and $\Theta(x)$ is the Heaviside step function.
After performing the integration, the average density matrices take the form of Eqs. \eqref{rhostat} and \eqref{rhortn} for the static noise and the RTN respectively.

\end{document}